\documentstyle[12pt]{article}
\begin{document}
\begin{titlepage}

\vspace*{2cm}
\begin{center}
{\large\bf Isgur-Wise Function and Observables of $\Lambda_b$ Baryon\\}
\end{center}

\vspace*{1cm}
\begin{center}
{\large\bf M.A.Ivanov, V.E.Lyubovitskij \\}
{\it Joint Institute for Nuclear Research\\}

\vspace*{1cm}
{\it 141980 Dubna, Moscow Region, Russia}\\
{\it e.mail:\hspace*{.2cm}ivanovm@thsun1.jinr.dubna.su,
\hspace*{.2cm} lubovit@thsun1.jinr.dubna.su}\\
\end{center}

\vspace*{4cm}
\begin{abstract}
In connection with planned experiments devoted to investigation
of semileptonic decays of beauty baryons Isgur-Wise function and
observables of $\Lambda_b$ baryon (decay rates and distributions, leptonic
spectra and asymmetry parameters) are calculated in the framework of diquark
model with taking into account of infrared regime for heavy quark.
\end{abstract}

\vspace*{2cm}


\end{titlepage}

\baselineskip 20pt

\section{Introduction}

Weak decays of heavy hadrons containing a single heavy quark
can be considered as a unique tool to determine of
Cabibbo-Kabayashi-Maskava matrix elements, original source to probe
hadronic structure and investigation of effects beyond Standard Model.

The growth of interest to such processes
is connected with considerable progress in experimental investigation
and new theoretical ideas which were borned. One has to remark that
during a long time the experimental programs were concentrated
mainly on heavy mesons and charm baryons.
Nowadays the facilities of modern accelerators allow to us investigate
properties of beauty baryons. Particular first observation of
$\Lambda_b$ baryon was made in the decay $\Lambda_b\to J/\Psi\Lambda$
at the $p\bar p$ collider (CERN) \cite{a1}. Semileptonic decays of
$\Lambda_b$ baryon $\Lambda_b\to\Lambda_c Xe\nu$ were studied by ALEPH
and OPAL group at LEP \cite{a2,a3}. Observation of $\Lambda_b$ production
in the $Z^0$ decays and measuring of $\Lambda_b$ lifetime were fulfilled
by DELPHI Collaboration \cite{a4}.

 From a theoretical point of view weak decays of heavy quarks
are interested because of a new type of symmetry of strong interactions
was discovered - spin-flavour symmetry in the world of heavy quarks
(Isgur-Wise symmetry) \cite{a5} and also heavy quark effective theory
(HQET) \cite{a5}-\cite{a7} - perturbative calculating scheme for
investigation properties of hadrons containing a single heavy quark.
Isgur-Wise symmetry manifested itself in the limit when heavy quark masses
go to infinity $m_Q\to\infty, Q=b,c$ (Isgur-Wise limit).
The consequences of these symmetries for weak heavy-hadron form factors
were worked out by Isgur and Wise in ref. \cite{a5,a6}.
Isgur and Wise showed \cite{a5}, that form factors of B-meson weak decays
$B\to D\ell\nu$ и $B\to D^\star\ell\nu$ are expressed through universal
function $\xi(\omega)$ (mesonic Isgur-Wise function),
where $\omega$ is the scalar product of the four-velocities of the initial
and final heavy-hadron states.

In paper \cite{a6} the consequences of Isgur-Wise symmetry
were obtained for form factors of semileptonic decays of heavy baryons.
It was shown that in the Isgur-Wise limit (baryonic form factors satisfied
to group conditions and expressed through three unknonw universal functions
$\zeta(\omega)$, $\eta(\omega)$ и $\iota(\omega)$ \cite{a6}.
For example, matrix element corresponding to decay
$\Lambda_b\to \Lambda_c e\nu$ in the Isgur-Wise limit has a trivial shape
$J_\mu(v,v^\prime)\propto
\zeta(\omega)\bar u(v^\prime)\gamma_\mu(1+\gamma_5)u(v)$.
For decays $\Omega_b(\Sigma_b)\to\Omega_c(\Sigma_c)e\nu$
и $\Omega_b(\Sigma_b)\to\Omega_c^\star(\Sigma_с^\star) e\nu$ connections
between form factors are more complicated. However, all form factors
can be expressed through two functions $\eta(\omega)$ and $\iota(\omega)$.
Unfortunatelly, HQET as realistic theory of heavy flavoured hadrons
has no posibility  to calculate functions $\zeta (\omega)$, $\eta (\omega)$
and $\iota(\omega)$. So calculation of barionic Isgur-Wise functions
got dissemination within various phenomenological approaches:
Infinite Momentum Frame (IMF) models \cite{a8,a9}, QCD Sum Rules \cite{a10},
Quark Confinement Models \cite{a11} and so on.

In \cite{a8,a9} semileptonic decay $\Lambda_b\to \Lambda_c e\nu$
was considered in the framework of IMF model where it was assumed that
heavy baryon consists of a heavy quark and light spin-zero diquark system.
In ref. \cite{a9} Bauer-Stech-Wirbel type infinite momentum frame wave
functions for the heavy $\Lambda$-type baryons were used and in \cite{a8}
Drell-Yan ones. Isgur-Wise function $\zeta(\omega)$ \cite{a8,a9} and
various observables of $\Lambda_b\to \Lambda_c e\nu$ decay
(rates, spectra and asymmetry parameters) were computed.
Using QCD Sum Rules technique $\zeta(\omega)$ form factor was calculted
in paper \cite{a10}. In \cite{a11} semileptonic decays of heavy baryons
$\Lambda_b\to\Lambda_c e\nu$ and $\Sigma_b\to\Sigma_c e\nu$ are considered
in the framework of Quark Confinement Model \cite{a12} with taking into
account of so-called {\it quark-diquark approximation} \cite{a11}.
It was showed in \cite{a11,a12} that all form factors of
$\Lambda_b\to\Lambda_c e\nu$ and $\Sigma_b\to\Sigma_c e\nu$ decay are
expressed through a single unversal function $\Phi(\omega)$ which
is equal identically to Isgur-Wise function $\zeta$ and is given the
expression $\Phi(\omega)\equiv\zeta(\omega)=
\ln(\omega+\sqrt{\omega^2-1})(\omega^2-1)^{-1/2}$.
Later it was founded by Xu \cite{a13} that obtained results within QCM
\cite{a11} contradict to Bjorken sum rule for semileptonic
$\Omega_b$ decay which give the following restriction on fucntion
$\Phi(\omega)$: $\Phi^2(\omega)\leq 3/(1+2\omega^2)$. So form factors
obtained in the QCM are suffiiciently "hard".

In our paper we obtained suppression of baryonic Isgur-Wise functions
using for heavy quark propagator so-called "infrapropagator", which
describes behaviour of heavy quark near its mass-shell, i.e. in the limit
when heavy quark mass goes to infinity (Isgur-Wise limit). Recently,
in \cite{a14,a15} idea about infrared behaviour of heavy quark in the limit
$m_Q\to\infty$ was successfully used in calculation of mesonic Isgur-Wise
function $\xi(\omega)$. Then we use obtained Isgur-Wise function in the
calculation of observables of $\Lambda_b$ baryon (decay rates and
distributions, leptonic spectra and asymmetry parameters).

\section{Model}

At the consideration of $b\to c$ decays of beaty baryons we will use
{\it diquark factorization of baryon structure}, which is based
on a observation that light degrees of freedom in these processes
are manifested itself as spectator or diquark. Thus heavy baryon
can be presented as two-particle bound state of heavy quark an light
diquark. In the case of $\Lambda_b$ and $\Lambda_с$ baryons light
diquark $D^a$ must possesses by the following quantum numbers:
spin $J=0$, izospin $I=0$ and positive P-parity.
Hence quark-diquark current with quantum numbers of $\Lambda_b$ and
$\Lambda_с$ baryons has the form $J_Q(x)=Q^a(x)D^a(x), \,\, Q=c, b$.
The interaction Lagrangain describing transition of heavy baryon $\Lambda_Q$
into quark-diquark pair and {\it vice versa} is written as
$${\cal L}_Q(x)=g_{\Lambda_Q}\bar\Lambda_Q J_Q + эрм.сопр.$$

Full Lagrangian needed for describing of decay
$\Lambda_b\to\Lambda_c\ell\nu_\ell$ is given by the formula:
\begin{eqnarray}
{\cal L}_{full}=\sum\limits_Q{\cal L}_Q + {\cal L}_{weak} + эрм.сопр.,
\;\;\;\;
{\cal L}_{weak}=\frac{G_F}{\sqrt{2}}\ell^\mu\bar c^aV_{bc}O_\mu b^a
\end{eqnarray}
\noindent where $V_{bc}$ is Kabayashi-Maskava matrix element.

\vspace*{.5cm}
\unitlength=1.00mm
\special{em:linewidth 0.4pt}
\linethickness{0.4pt}
\begin{picture}(127.00,64.00)
\put(45.00,12.00){\circle*{4.00}}
\put(95.00,12.00){\circle*{4.00}}
\put(45.00,11.65){\line(2,0){50.00}}
\put(45.00,11.75){\line(2,0){50.00}}
\put(45.00,12.00){\line(2,0){50.00}}
\put(45.00,12.25){\line(2,0){50.00}}
\put(95.00,12.00){\line(1,0){25.00}}
\put(120.00,13.00){\line(-1,0){25.00}}
\put(95.00,11.00){\line(1,0){25.00}}
\put(45.00,11.00){\line(-1,0){25.00}}
\put(20.00,12.00){\line(1,0){25.00}}
\put(45.00,13.00){\line(-1,0){25.00}}
\put(45.00,12.00){\line(5,6){25.00}}
\put(70.00,42.00){\line(5,-6){25.00}}
\put(70.00,42.00){\line(0,1){3.00}}
\put(70.00,46.00){\line(0,1){3.00}}
\put(70.00,50.00){\line(0,1){3.00}}
\put(70.00,54.00){\line(0,1){3.00}}
\put(70.00,58.00){\line(0,1){3.00}}
\put(75.00,47.00){\line(0,1){12.00}}
\put(89.00,52.00){\makebox(0,0)[cc]{$q=p-p^\prime$}}
\put(70.00,64.00){\makebox(0,0)[cc]{$\ell\bar\nu_\ell$}}
\put(13.00,12.00){\makebox(0,0)[cc]{$\Lambda_c$}}
\put(127.00,12.00){\makebox(0,0)[cc]{$\Lambda_b$}}
\put(108.00,18.00){\makebox(0,0)[cc]{$p$}}
\put(32.00,18.00){\makebox(0,0)[cc]{$p^\prime$}}
\put(53.00,32.00){\makebox(0,0)[cc]{$p^\prime-k$}}
\put(88.00,32.00){\makebox(0,0)[cc]{$p-k$}}
\put(70.00,5.00){\makebox(0,0)[cc]{$k$}}
\put(32.00,12.00){\line(2,1){4.00}}
\put(32.00,12.00){\line(2,-1){4.00}}
\put(108.00,12.00){\line(2,1){4.00}}
\put(108.00,12.00){\line(2,-1){4.00}}
\put(75.00,59.00){\line(-1,-4){1.00}}
\put(75.00,59.00){\line(1,-4){1.00}}
\put(58.00,28.00){\line(1,2){2.00}}
\put(58.00,28.00){\line(2,1){4.00}}
\put(79.00,31.00){\line(2,-1){4.00}}
\put(79.00,31.00){\line(1,-3){1.30}}
\put(68.00,12.00){\line(5,2){5.00}}
\put(68.00,12.00){\line(5,-2){5.00}}
\put(45.00,7.00){\makebox(0,0)[cc]{$I$}}
\put(95.00,7.00){\makebox(0,0)[cc]{$I$}}
\put(66.00,42.00){\makebox(0,0)[cc]{$O_\mu$}}
\end{picture}

\vspace*{.5cm}

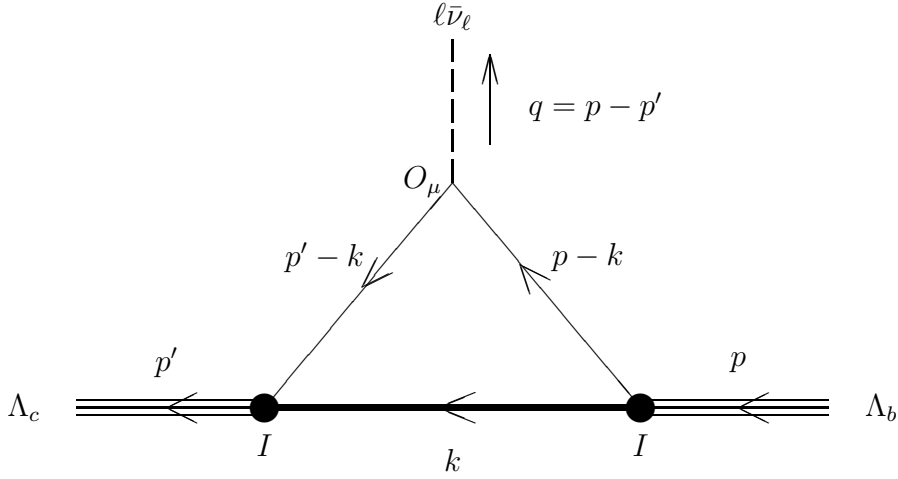
\hspace*{3.5cm}Fig.1 Semileptonic decay of $\Lambda_b$ baryon.

\newpage

The corresponding Feynman diagram is drawn in Fig.1.
As light diquark propagator we will use standard propagator of
scalar field $S_D(p^2)=1/(M^2_D-p^2)$ where $M_D$ is a diquark mass.

As it is known from HQET \cite{a5}-\cite{a7} in the Isgur-Wise limit
heavy quark is near its mass-shell, i.e. infrared regime comes for heavy
quark. Infrared asymptotics of one-particle Green function was investigated
in an Abelian theory (Quantum Electrodynamics) in many papers
(see, for example, ref. \cite{a16}). So-called {\it infrapropagator}
of electron has the following shape
\begin{equation}
G(p,\nu)=(m-\not\! p)^{-1}s(p^2,\nu), \;\;\;
s(p^2,\nu)= (1-p^2/m^2)^{-\nu}
\end{equation}
\noindent Here parameter $\nu$ is connected with gauge parameter $d_\ell$
by well-known condition
\begin{equation}
\nu=(\alpha_{em}/4\pi)(3-d_\ell),
\end{equation}
where $\alpha_{em}$ is a fine structure constant.

In the papers \cite{a14,a15} as the first approximation
{\it infrapropagator} of an Abelian theory is used in the calculations of
Isgur-Wise function. We also will use this propagator.
Particular heavy quark propagator is given by formula
\begin{equation}
S_Q(p,\nu)=\frac{m_Q+ \not\! p}{m_Q^2}
\biggl(\frac{1}{1-p^2/m_Q^2}\biggr)^{1+\nu}
\end{equation}
The parameter $\nu$ must be bigger than zero to make all matrix elements
ultroviolet finite. Therefore our model contains three parameters:
light diquark mass $M_D$, parameter $\bar\Lambda=M_{B_Q}-m_{B_Q}$ and
infrared parameter $\nu$. Experimental restrictions on the parameter
$\bar\Lambda$ are absent. Theoretical evaluation of value $\bar\Lambda$
was made using QCD Sum Rules Technique by Neubert \cite{a17}.
It was founded that $\bar\Lambda=0.50\pm 0.07$ GeV.
Within QCM under consideration of leptonic decays of B and D mesons
it was obtained that the parameter $\bar\Lambda$ must exchange in limits
$0\leq\bar\Lambda\leq 0.6$ GeV. In our calculations we will fix the
parameter $\bar\Lambda$ equaled to 0.6 GeV (as in various phenomenological
approaches, see for example ref. \cite{a9}).

\section{$\Lambda_b\to\Lambda_c$ Isgur-Wise Function}

Isgur-Wise function in our model depends on parameter $\nu$ and ratio of
diquark mass and parameter $\Lambda$: $R=M_D/\bar\Lambda$ and is written as
\begin{eqnarray}
\zeta(\omega,R,\nu)=
\frac{\Phi(\omega,R,\nu)}{\Phi(1,R,\nu)}, \;\;
\Phi(\omega,R,\nu)=
\int\limits_0^\infty duu^{1+2\nu}\int\limits_0^1 \frac{dt}{\sqrt{1-t}}
\frac{t^\nu}{[R^2+(u-1)^2+\frac{u^2t}{2}(\omega-1)]^{1+2\nu}}
\nonumber
\end{eqnarray}
In our calculations of Isgur-Wise function we try to get the best agreement
with results of IMF models \cite{a8,a9}. It was achieved when parameter
$\nu$ is closer to value 1. Results for $\omega$-dependence of
Isgur-Wise function for parameter $\nu=1$ and various meaning of parameter
$R$ in the interval $1.2<R<2$ are presented in Fig.2. For comparison
results of other theoretical approaches are given (IMF model \cite{a9}
and QCD Sum Rules \cite{a10}). One can see that a good agreement with IMF
model is achieved when $R=1.25$. This choice of parameter $R$ will called
in our paper as {\it the best fit}. It is clear that in this case the diquark
mass is equal to 0.7 GeV.

Let us to calculate also charge radius of Isgur-Wise function
$\rho^2=-d\zeta/d\omega|_\omega=1$
\begin{eqnarray}
\rho^2=
\frac{(1+\nu)(1+2\nu)}{3+2\nu}\frac
{\int\limits_0^\infty duu^{3+2\nu}[R^2+(u-1)^2]^{-2-2\nu}}
{\int\limits_0^\infty duu^{1+2\nu}[R^2+(u-1)^2]^{-1-2\nu}}
\end{eqnarray}
The results for $\rho^2$ as function of $\bar\Lambda$ are given in Table 1.

\vspace*{1cm}

\hspace*{2.5cm}Table 1. Charge radius of Isgur-Wise function.
\def\arraystretch{2.0}
\begin{center}
\begin{tabular}{|c|c|c|c|c|c|c|c|}
\hline\hline
$\bar\Lambda$ (GeV)& 0.47 & 0.48 & 0.49 & 0.50 & 0.51 & 0.52 & 0.53
\\[0.5cm]
\hline
$\rho^2$ & 3.03 & 2.67 & 2.43 & 2.25 & 2.10 & 1.99 & 1.89
\\[0.5cm]
\hline\hline
\end{tabular}
\end{center}

\section{Weak Properties of $\Lambda_b$ Baryon}

Observables of semileptonic decays of $\Lambda_b$ baryon
(decay rates, differential distributions, leptonic spectra and
asymmetry parameters) we will determine in the terms of helicity amplitudes
$H^\Gamma_{\lambda_f\lambda_W}$ \cite{a9,a18}, where $\lambda_f$ is
helicity of baryon in the final state and $\lambda_W$ is helicity of W-boson
out of mass-shell.
We will present the calculations of $\Lambda_b$ baryon
properties in the case of {\it the best fit}
In the Isgur-Wise limit helicity amplitudes are expressed
through function $\zeta(\omega)$:
\begin{eqnarray}
& &H^V_{\pm\frac{1}{2}0}=\zeta\sqrt{\frac{\omega-1}{\omega_{max}-\omega}}
(M_{\Lambda_b}+M_{\Lambda_c}),
\;\;
H^A_{\pm\frac{1}{2}0}=\pm\zeta\sqrt{\frac{\omega+1}{\omega_{max}+\omega}}
(M_{\Lambda_b}-M_{\Lambda_c}),\nonumber \\
& &H^V_{\pm\frac{1}{2}1}=-2\zeta\sqrt{M_{\Lambda_b}M_{\Lambda_c}(\omega-1)},
\;\;
H^A_{\pm\frac{1}{2}1}=\mp 2\zeta\sqrt{M_{\Lambda_b}M_{\Lambda_c}(\omega+1)},
\;\;
\omega_{max}=\frac{M^2_{\Lambda_b}+M^2_{\Lambda_c}}
{2M_{\Lambda_b}M_{\Lambda_c}}
\nonumber
\end{eqnarray}

\noindent Decay rates of semileptonic decays are calculated in accordance
with formula
\begin{eqnarray}
\Gamma=\int\limits_1^{\omega_{max}}d\omega \;\;\frac{d\Gamma}{d\omega},
\;\;\;\frac{d\Gamma}{d\omega}=\frac{d\Gamma_{T_+}}{d\omega}+
\frac{d\Gamma_{T_-}}{d\omega}+\frac{d\Gamma_{L_+}}{d\omega}+
\frac{d\Gamma_{L_-}}{d\omega}
\end{eqnarray}
\noindent where indices $T$ and $L$ denote partial contributions of
the transverse $(\lambda_W=\pm 1)$ and longitudinal $(\lambda_W=0)$ components
of the current transition.
Partial differential distributions are equal
\begin{eqnarray}
\frac{d\Gamma_{T_\pm}}{d\omega}=\kappa_\omega |H_{\pm\frac{1}{2}\pm 1}|^2, \;\;
\frac{d\Gamma_{L_\pm}}{d\omega}=\kappa_\omega |H_{\pm\frac{1}{2}0}|^2, \;\;
\kappa_\omega=
\frac{G_F^2}{(2\pi)^3}|V_{bc}|^2
\frac{M^3_{\Lambda_c}}{6}(\omega_{max}-\omega)\sqrt{\omega^2-1}
\end{eqnarray}
\noindent where $H_{\lambda_f\lambda_W}=
H^V_{\lambda_f\lambda_W}-H^A_{\lambda_f\lambda_W}$.
The results for decay rates are given in Table 2. For comparison
we present the results of IFM model \cite{a9}. The curves of differential
distributions are drawn in Fig.3. One can to underline that our results
for $\frac{d\Gamma}{d\omega}$ are practically coincide with ones obtained
in ref. \cite{a9}.

Leptonic spectra $d\Gamma/dE_\ell$ is calculated by formula
\begin{eqnarray}
\frac{d\Gamma}{dE_\ell}=\frac{d\Gamma_{T_+}}{dE_\ell}
+\frac{d\Gamma_{T_-}}{dE_\ell}+\frac{d\Gamma_{L_+}}{dE_\ell}
+\frac{d\Gamma_{L_-}}{dE_\ell}
\end{eqnarray}
\noindent Expressions for partial leptonic spectra are given by formulas
\begin{eqnarray}
& &\frac{d\Gamma_{T_\pm}}{dE_\ell}=\hspace*{-0.5cm}\int\limits_{\omega_{min}
(E_\ell)}^{\omega_{max}}\hspace*{-0.5cm}d\omega\hspace*{0.1cm}
\kappa_E(1\pm \cos\Theta)^2
|H_{\pm\frac{1}{2}\pm 1}|^2\;\;\;
\frac{d\Gamma_{L_\pm}}{dE_\ell}=\hspace*{-0.5cm}\int\limits_{\omega_{min}
(E_\ell)}^{\omega_{max}}\hspace*{-0.5cm}d\omega\hspace*{0.1cm}
\kappa_E(1-\cos^2\Theta)^2
|H_{\pm\frac{1}{2}0}|^2, \nonumber\\
&\mbox{где}&\kappa_E=\frac{G_F^2}{(2\pi)^3}|V_{bc}|^2
\frac{M^2_{\Lambda_c}}{8}(\omega_{max}-\omega),
\;\;\;
\cos\Theta=\frac{E_\ell^{max}-2E_\ell+M_{\Lambda_c}(\omega_{max}-\omega)}
{M_{\Lambda_c}\sqrt{\omega^2-1}}, \nonumber \\
& &E^{max}_\ell=\frac{M^2_{\Lambda_b}-M^2_{\Lambda_c}}{2M_{\Lambda_b}},
\;\;\;
\omega_{min}(E_\ell)=\omega_{max}-2\frac{E_\ell}{M_{\Lambda_c}}
\frac{E^{max}_\ell-E_\ell}{M_{\Lambda_b}-2E_\ell}
\nonumber
\end{eqnarray}
\noindent Our results for leptonic spectra are pictured on Fig.4.

\vspace*{1cm}

\hspace*{2.5cm}Table 2. Decay rate of
$\Lambda_b\to\Lambda_c\ell\nu_\ell$ decay (in units $10^{10}$ sec$^{-1}$)
\def\arraystretch{3.0}
\begin{center}
\begin{tabular}{|c|c|c|c|c|c|c|c|}
\hline\hline
Model & $\Gamma_{total}$ & $\Gamma_T$ & $\Gamma_{T_+}$ & $\Gamma_{T_-}$ &
$\Gamma_L$ & $\Gamma_{L_+}$ & $\Gamma_{L_-}$ \\
\hline
QCM & 4.07 & 1.74 & 0.47 & 1.27 & 2.32 & 0.11 & 2.21\\
\hline
IMF \cite{a9} & 4.57 & 1.88 & 0.42 & 1.46 & 2.69 & 0.11 & 2.58\\
\hline\hline
\end{tabular}
\end{center}

Let us to consider two-cascade weak decay
$\Lambda_b\to\Lambda_c[\to \Lambda_s\pi]+W[\to\ell\nu_\ell]$ which is
characterizes by asymmetry parameters. Formalism and detailed analysis of
asymmetry parameters was fulfiled in paper \cite{a14}. Here we calculate
asymmetry parameters of nonpolarized $\Lambda_b$ decay
($\alpha, \alpha^\prime, \alpha^{\prime\prime},
\gamma$) and polarized one ($\alpha_P, \gamma_P$) which in the terms of
helicity amplitudes are given by following expressions
\begin{eqnarray}
& &\alpha=\frac{H^-_T+H^-_L}{H^+_T+H^+_L},\;\;\;
\alpha^\prime=\frac{H^-_T}{H^+_T+2H^+_L},\;\;\;
\alpha^{\prime\prime}=\frac{H^+_T-2H^+_L}{H^+_T+2H^+_L},\;\;\;
\gamma=\frac{2H_\gamma}{H^+_T+H^+_L},\nonumber\\
& &\alpha_P=\frac{H^-_T-H^-_L}{H^+_T+H^+_L},
\gamma_P=\frac{2H_{\gamma_P}}{H^+_T+H^+_L},\\
\mbox{где}& &H^\pm_T=|H_{1/2\;1}|^2\pm|H_{-1/2\;-1}|^2
\;\;\;
H^\pm_L=|H_{1/2\;0}|^2\pm|H_{-1/2\;0}|^2\nonumber\\
& &H_\gamma=Re(H_{-1/2\;0}H^*_{1/2\;1}+H_{1/2\;0}H^*_{-1/2\;-1})\;\;\;
H_{\gamma_P}=Re(H_{1/2\;0}H^*_{-1/2\;0})
\nonumber
\end{eqnarray}

\noindent In our paper we will calculate average meanings of asymmetry
parameters ($<\alpha>, <\alpha^\prime>$ and so on) as result of separate
integration over parameter $\omega$ of numerator and denomerator of eq. (8)
with weight $(\omega_{max}-\omega)\sqrt{\omega^2-1}$ in the interval
$1\leq \omega\leq \omega_{max}$.
Results for average meanings are given in Table 3. Also the results of paper
\cite{a9} are presented.

\vspace*{1cm}

\hspace*{2.5cm}Table 3. Asymmetry parameters
\def\arraystretch{3.0}
\begin{center}
\begin{tabular}{|c|c|c|c|c|c|c|}
\hline\hline
Model & $<\alpha>$ & $<\alpha^\prime>$ & $<\alpha^{\prime\prime}>$ &
$<\gamma>$ & $<\alpha_P>$ & $<\gamma_P>$ \\
\hline
QCM & -0.71 & -0.13 & -0.46 & 0.61 & 0.32 & -0.19\\
\hline
IMF \cite{a9}& -0.71 & -0.12 & -0.46 & 0.61 & 0.33 & -0.19\\
\hline\hline
\end{tabular}
\end{center}

\noindent Paper is supported in part by the Russian Fond of Fundamental
Research (RFFR) under contract 94-02-03463-a.


\newpage

\begin{thebibliography}\\

\bibitem{a1} UA1 Collab., Albajar C., et al., Phys.Lett., {\bf B273}, 540
(1991).
\bibitem{a2} ALEPH Collab., Buskulic D., et al., Phys.Lett., {\bf B294}, 145
(1992).
\bibitem{a3} OPAL Collab., Acton P.D., et al., Phys.Lett., {\bf B281}, 394
(1992).
\bibitem{a4} DELPHI Collab., Abreu P., et al., Phys.Lett., {\bf B311}, 379
(1993).
\bibitem{a5} Isgur N., Wise M., Phys.Lett., {\bf B232}, 113 (1989); {\bf B237},
527 (1990).
\bibitem{a6} Isgur N, Wise M., Nucl.Phys., {\bf B348}, 276 (1991).
\bibitem{a7} Georgi H., Phys.Lett., {\bf B240}, 447 (1990).
\bibitem{a8} Guo X.-H., Kroll P.,  Z.Phys., {\bf C59}, 567 (1993).
\bibitem{a9} K\"{o}nig, K\"{o}rner J.G., et al., Preprint {\bf DESY 93-011},
1993.
\bibitem{a10} Grozin A.G. and Yakovlev O.I., Phys.Lett., {\bf B291}, 441
(1992).
\bibitem{a11} Efimov G.V., Ivanov M.A., et. al.,  Z.Phys., {\bf C54}, 349
(1992).
\bibitem{a12} Efimov G.V., et al. The Quark Confinement Model of
 Hadrons.- IOP Publishing, Bristol $\&$ Philadelphia, 1993.
\bibitem{a13} Xu Q.P., Phys.Rev., {\bf D48}, 5429 (1993).
\bibitem{a14} Karanikas A.I., Ktorides C.N. and Stefanis N.G., Phys.Lett., {\bf
B301}, 397 (1993).
\bibitem{a15} Ivanov M.A., Mizutani T., Preprint {\bf hep-ph/9406226}.
\bibitem{a16} N.N.Bogolubov, D.V.Shirkov, Introduction to the
 Theory of Quantized Fields, Interscience Publishers Inc.,
 New York, 1959.
\bibitem{a17} Neubert M., Phys.Rev., {\bf D45}, 2451 (1992).
\bibitem{a18} K\"{o}rner J.G. and Kr\"{a}mer M., Phys.Lett., {\bf B275}, 495
(1992).
\end{thebibliography}
\end{document}